\documentclass{aa}
\usepackage[dvips]{graphicx}
\usepackage{psfig}
\begin{document}
   \title{New neighbours: IV. 30 DENIS late-M dwarfs \\
          between 15 and 30 parsecs}


   \author{N.~Phan-Bao
       \inst{1,2}
       \and
       J.~Guibert
       \inst{1,2}
       \and
       F.~Crifo
       \inst{2}
       \and
       X.~Delfosse
       \inst{3}
       \and
       T.~Forveille
       \inst{3,4}
       \and
       J.~Borsenberger
       \inst{5}
       \and
       N.~Epchtein
       \inst{6}
       \and
       P.~Fouqu\'e
       \inst{7,8}
       \and
       G.~Simon
       \inst{2}
       }

   \offprints{N. Phan-Bao, \email{bn.phan@obspm.fr}}

   \institute{Centre d'Analyse des Images, DASGAL, Observatoire de Paris,  61 avenue 
              de l'Observatoire, 75014 Paris, France
             \and
              Observatoire de Paris (DASGAL/UMR-8633), 75014 Paris, France
             \and
              Laboratoire d'Astrophysique de Grenoble, Universit\'e J. 
              Fourier, B.P. 53, F-38041 Grenoble, France
             \and
              Canada-France-Hawaii Telescope Corporation, 65-1238 Mamalahoa 
              Highway, Kamuela, HI 96743 USA
             \and
              Institut d'Astrophysique de Paris, 98bis boulevard Arago, 
              75014 Paris
             \and
              Observatoire de Nice, BP 4229, 06304 Nice Cedex 4, France
             \and
              DESPA, Observatoire de Paris, 5 place J. Janssen, 92195 
              Meudon Cedex, France
             \and
              European Southern Observatory, Casilla 19001, Santiago 19, Chile
             }

   \date{Received / Accepted}

\abstract{
We present 30 new nearby (d~$<$~30 pc) red dwarf candidates, with
estimated spectral types M6 to M8. 26 were directly selected from the 
DENIS database, and another 4 were first extracted from the LHS catalogue
and cross-identified with a DENIS counterpart. Their proper motions were
measured on the MAMA measuring machine from plates spanning 13 to 48~years,
and are larger than 0.1\arcsec yr$^{\rm -1}$, ruling out the possibility that
they are giants. Their distances were estimated from 
the DENIS colours and IR colour-magnitude relations and range
between 15 and 30~pc.
}

\titlerunning{New neighbours: IV. 30 DENIS late-M dwarfs}
\authorrunning{N.Phan-Bao et al.}
\maketitle

\section{Introduction}

Much of our understanding of stellar astronomy rests upon the nearest
stars. As individual objects they are the brightest and hence best
studied examples of their spectral type, and their trigonometric parallaxes
can be measured accurately, although with significant effort.
The solar neighbourhood sample also provides deep insight into the nature 
of our Galaxy's components, 
through studies of its stellar luminosity and mass functions, its 
kinematics, chemical composition, and multiplicity statistics. 

Perhaps surprisingly, this sample is still incomplete, even very close to 
the Sun, as illustrated by the recent discoveries of three
new stars with d~$<$~6~pc (Henry et al. \cite{henry}; Delfosse et al. 
\cite{delfossec}; Scholz et al. \cite{scholz01} ).
>From a comparison of the observed star densities within 5 and 
10~parsecs, Henry et al. (\cite{henry}) estimate that approximately
130~systems are missing from the 10~pc sample. Most of these missing stars
are red M dwarfs, with B$-$V~$>$~1.70 (Gliese et al. \cite{gliese}),
and the deficit is largest south of declination $-30\degr$.
 
For the most part the known members of the solar neighbourhood have
been selected within the available catalogues of
trigonometric parallaxes, though some significant fraction is still included
on the basis of photometric or spectroscopic distances (Gliese \cite{gliese56};
Gliese \& Jahrei\ss ~\cite{gliese79}, \cite{gliese91}). However, for at least the
last fifty years, parallax programs have selected their targets for a good part
from proper motion catalogues,
such as the Cincinnati catalogues (Porter et al. \cite{porter18}, \cite{porter30}),
the Luyten catalogues (LFT \cite{luyten55}; LTT \cite{luyten57}, \cite{luyten61};
LHS \cite{luytena}; NLTT \cite{luytenb})
or the Lowell catalogues (Giclas et al. \cite{giclas71}, \cite{giclas78}).
This leads to a proper motion bias of the
resulting sample, at least for the faint end. 
To date even the smaller LHS catalogue has been incompletely
followed-up, and the Luyten catalogues therefore contain many unrecognized 
nearby stars (e.g. Scholz et al. \cite{scholz01}, for an extreme example).
Several groups are working to identify them (Henry et al. \cite{henry}; 
Gizis \& Reid \cite{gizis97}; Jahrei\ss~et al. \cite{jahreiss01}). 
One limitation to these efforts is the much brighter
limiting magnitude of the proper motion catalogues south of declination $-33\degr$
(e.g. Scholz et al. \cite{scholz01}, for a 
recent discussion), where a much larger fraction of the solar 
neighbourhood stars is currently missing.

The recent near-infrared sky surveys, DENIS and 2MASS, represent 
a powerful alternative tool to identify these faint and cool nearby 
stars. Candidates are first selected on simple colour and photometric 
distance criteria, producing an initial list that also contains many giants, 
and potentially some pre-main-sequence stars. At low galactic 
latitudes it would also include many reddened distant stars.
In a second step these contaminating populations need to be
rejected, through either (1) a systematic spectroscopic follow up 
(Gizis et al. \cite{gizis}), or (2) proper motion selection (distant 
giants have very small proper motions), or (3) more accurate multi-colour
photometry than
available from the surveys. Systematic spectroscopy is clearly 
the cleanest approach, and does not reject the few nearby star that, like
Gl~710 for instance, have small proper motions. Proper motion selection 
on the other hand is cheap and effective, but it obviously does nothing 
to correct for the proper motion bias of the present nearby star catalogues.

We are using here the DENIS near infrared sky survey to identify
possible very-low-mass stars within 30 parsecs. We conservatively adopt 
a distance cutoff slightly larger than the 25~pc limit of the Catalogue of 
Nearby Stars of Gliese and Jahrei\ss~(CNS3, \cite{gliese91}), 
to avoid losing bona-fide d~$<$~25~pc stars from imprecise photometric
distances.
One initial result of this effort was the discovery of an M9 dwarf 
at a spectroscopic distance of only 4~pc, DENIS-P~J104814.7-395606.1 
(Delfosse et al. \cite{delfossec}). In this paper we present 26 nearby 
red dwarf candidates, selected with I$-$J~$>$~2.0 (M6 or later) in the DENIS 
database and with a proper motion larger than 0.1\arcsec yr$^{\rm -1}$.
13 of these are new, and another 13 were known as high-PM objects but 
had not been characterized further. Four additional candidates were
first selected from the southern part of the LHS catalogue
and then identified with DENIS counterparts.

Sect.~2 reviews the sample selection, and Sect.~3 presents the 
photographic photometry and the proper motion measurements. 
We discuss the results in Sect.~4 and summarize in Sect.~5.
\section{ Sample selection}
\subsection{DENIS observations}
\begin{table}
\begin{tabular}{llll} 
\hline
\hline
\noalign{\smallskip}
I$-$J & Spectral type & M$_{\rm I}$ & D$_{\rm lim}$ \\ 
\hline
\noalign{\smallskip}
2.0 & M5.5 - M6 & 11.5 & 250~pc \\
2.5 & M6.5 - M7 & 12.9 & 130~pc \\
3.0 & M8  & 14.0 & 80~pc \\ 
\noalign{\smallskip}
\hline
\end{tabular}
\caption{Typical detection limiting distances
(D$_{\rm lim}$) 
for M dwarfs with
colour 2.0 ${\leq}$ I$-$J ${\leq}$ 3.0 in the DENIS survey. The spectral 
types are from Leggett
(\cite{leggett}) and the absolute magnitude is taken from 
Figure~\ref{fig_col_mag}.
}
\label{table_dist}
\end{table}
\begin{figure}
\psfig{height=6.5cm,file=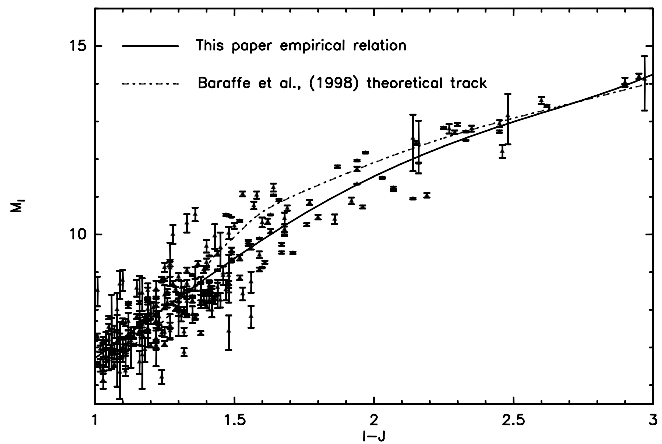,angle=-90} 
\caption{$M_I : I-J$ HR diagram for M dwarfs with known distances (from
  Leggett \cite{leggett} and Tinney et al. \cite{tinney}). Our
  empirical polynomial fit and the 3~Gyr model of Baraffe et al. (\cite{baraffe98}) for a solar
  metalicity are overlaid.}
\label{fig_col_mag}
\end{figure}
The DEep Near-Infrared Survey (DENIS) is a southern sky 
survey (Epchtein \cite{epchtein97}), which will provide
full coverage of the southern hemisphere in two near-infrared bands (J
and K$_s$) and one optical band (I).
DENIS observations are carried out on the ESO 1m telescope at La Silla.
Dichroic beam splitters separate the three channels, and a
focal reducing optics provide scales of 3\arcsec~per pixel on the
256$\times$256 NICMOS3 arrays used for the two infrared channels, and
1\arcsec~per pixel on the 1024$\times$1024 Tektronix CCD detector of the I
channel. 

The image data were processed with the
standard DENIS software pipeline (Borsenberger \cite{borsen97}, 
Borsenberger et al., in preparation) at the 
Paris Data Analysis Center (PDAC).  The thermal background produced
by the instrument and the sky emission is derived from a local clipped 
mean along the
strip. Flat-field corrections are derived from observation of the
sunrise sky.
Source extraction and photometry are performed at PDAC, using 
a space-varying kernel algorithm (Alard \cite{alard00}). 
The astrometry of the individual DENIS frames is referenced to the 
USNO-A2.0 catalogue, whose $\sim$~1\arcsec~accuracy therefore determines 
the absolute precision of the DENIS positions.

With a 100\% completeness level at I$\sim$18.0, J$\sim$16.0 and
K$_s\sim$13.0, the DENIS survey is sensitive to M
dwarfs out to large distances (Table \ref{table_dist}),
well beyond the 25~pc limit of the CNS3 (\cite{gliese91}).
In this part of the HR diagram the I$-$J (or alternately I$-$K) 
colour index is an excellent spectral type and luminosity estimator 
(Leggett \cite{leggett}). The DENIS data therefore provide
immediate identification of M dwarf candidates from I$-$J~$>$~1.0, and, 
under the assumption that they are indeed dwarfs, a good estimate of their 
distance through colour-magnitude relations. 
\begin{table*}
   \caption{Position and observation information for the 30 new nearby
    star candidates}
    \label{obs}
  $$
   \begin{tabular}{lllllll}
   \hline 
   \hline
   \noalign{\smallskip}
DENIS Name  & Other name & $\alpha_{\rm 2000}$ & $\delta_{\rm 2000}$ & DENIS    & Number of    & Time  \\
            &            &                 &                 & Epoch    & observations & baseline [yr] \\
   \hline
   \noalign{\smallskip}
DENIS-P J0004575$-$170937 &...      &00 04 57.54& $-$17 09 37.0&1999.521&~\,~\,~\,4 & 47.834  \\
DENIS-P J0013466$-$045736 &LHS 1042 &00 13 46.60& $-$04 57 36.5&1999.846&~\,~\,~\,4 & 45.173  \\        
DENIS-P J0019275$-$362015 &...      &00 19 27.53& $-$36 20 15.7&1999.877&~\,~\,~\,3 & 21.064  \\
DENIS-P J0041353$-$562112 &...      &00 41 35.39& $-$56 21 12.9&1999.803&~\,~\,~\,3 & 20.984  \\
DENIS-P J0145434$-$372959 &...      &01 45 43.49& $-$37 29 59.3&1999.877&~\,~\,~\,3 & 22.165  \\
DENIS-P J0253444$-$795913 &...      &02 53 44.41& $-$79 59 13.6&1999.699&~\,~\,~\,3 & 21.921  \\
DENIS-P J0312251$+$002158 &...      &03 12 25.11& $+$00 21 58.6&1999.704&~\,~\,~\,3 & 48.020  \\
DENIS-P J0324268$-$772705 &...      &03 24 26.88& $-$77 27 05.5&1999.855&~\,~\,~\,3 & 22.077  \\ 
DENIS-P J0413398$-$270428 &LP 890- 2&04 13 39.81& $-$27 04 28.9&1999.838&~\,~\,~\,4 & 41.869  \\
DENIS-P J0602542$-$091503$^{+}$&LHS 1810 &06 02 54.26& $-$09 15 03.8&1999.929&~\,~\,~\,3 & 17.040  \\
DENIS-P J0848189$-$201911$^{+}$&LHS 2049 &08 48 18.94& $-$20 19 11.4&1999.258&~\,~\,~\,3 & 21.014  \\
DENIS-P J1003191$-$010507$^{+}$&LHS 5165 &10 03 19.18& $-$01 05 07.7&1999.030&~\,~\,~\,3 & 13.011  \\
DENIS-P J1136409$-$075511 &LP 673-63&11 36 40.99& $-$07 55 11.7&1999.373&~\,~\,~\,4 & 45.195  \\
DENIS-P J1216101$-$112609 &LP 734-87&12 16 10.16& $-$11 26 09.9&1999.203&~\,~\,~\,4 & 44.948  \\
DENIS-P J1223562$-$275746 &LHS 325a &12 23 56.27& $-$27 57 46.7&1999.304&~\,~\,~\,4 & 41.162  \\
DENIS-P J1236153$-$310646 &LP 909-55&12 36 15.32& $-$31 06 46.0&1999.249&~\,~\,~\,4 & 41.923  \\
DENIS-P J1357149$-$143852 &...      &13 57 14.97& $-$14 38 52.6&1999.225&~\,~\,~\,4 & 44.959  \\
DENIS-P J1406493$-$301828$^{+}$&LHS 2859 &14 06 49.33& $-$30 18 28.0&1999.395&~\,~\,~\,3 & 24.927  \\
DENIS-P J1412069$-$041348 &LP 679-32&14 12 06.98& $-$04 13 48.2&1999.444&~\,~\,~\,4 & 42.118  \\
DENIS-P J1553251$-$044741 &...      &15 53 25.13& $-$04 47 41.3&1999.581&~\,~\,~\,4 & 45.170  \\ 
DENIS-P J1614252$-$025100 &LP 624-54&16 14 25.20& $-$02 51 00.5&1999.280&~\,~\,~\,4 & 45.763  \\ 
DENIS-P J2002134$-$542555 &...      &20 02 13.42& $-$54 25 55.7&1999.581&~\,~\,~\,4 & 23.077  \\ 
DENIS-P J2049527$-$171608 &LP 816-10&20 49 52.72& $-$17 16 08.6&1999.518&~\,~\,~\,3 & 44.946  \\ 
DENIS-P J2107247$-$335733 &...      &21 07 24.73& $-$33 57 33.3&1999.581&~\,~\,~\,4 & 19.812  \\ 
DENIS-P J2134222$-$431610 &WT 792   &21 34 22.27& $-$43 16 10.4&1999.767&~\,~\,~\,3 & 21.189  \\
DENIS-P J2202112$-$110945*&LP 759-17&22 02 11.28& $-$11 09 45.8&1999.419&~\,~\,~\,4 & 45.672  \\
DENIS-P J2213504$-$634210 &WT 887   &22 13 50.47& $-$63 42 10.0&1999.537&~\,~\,~\,3 & 22.976  \\
DENIS-P J2331217$-$274949 &...      &23 31 21.73& $-$27 49 49.9&1999.869&~\,~\,~\,5 & 45.119  \\ 
DENIS-P J2333405$-$213353 &LHS 3970 &23 33 40.59& $-$21 33 53.2&1999.770&~\,~\,~\,4 & 45.176  \\ 
DENIS-P J2353594$-$083331 &...      &23 53 59.44& $-$08 33 31.6&1999.693&~\,~\,~\,3 & 46.077  \\ 
    \noalign{\smallskip}
    \hline 
   \end{tabular}
  $$
  \begin{list}{}{}
  \item[$^{+}$] Objects selected from LHS initially
  \item[$^{*}$] Already listed by Gizis et al. (\cite{gizis})

Columns 1 \& 2: Object name in the DENIS data base and other \
identification if available. 

Columns 3, 4 \& 5: DENIS Position with respect to equinox J2000 at DENIS epoch. 

Columns 6 \& 7: Number of position observations and time baseline. 
  \end{list}
\end{table*}
\begin{table*}
   \caption{34 red DENIS stars with $\mu < 0.1$\arcsec yr$^{\rm -1}$, not kept for further
   investigation in the present paper}
    \label{obs1}
  $$
   \begin{tabular}{lllllll}
   \hline 
   \hline
   \noalign{\smallskip}
DENIS Name                & $\alpha_{\rm 2000}$ & $\delta_{\rm 2000}$      &  DENIS & I   &  I$-$J  &  J$-$K  \\
                          &                 &                      &  Epoch  &     &         &         \\
   \hline
   \noalign{\smallskip}
DENIS-P J0013093$-$002551 & 00 13 09.34     & $-$00 25 51.5  & 1999.838 &  14.37  &  2.22   &  0.88    \\
DENIS-P J0100021$-$615627 & 01 00 02.13     & $-$61 56 27.1  & 1999.964 &  15.01  &  2.42   &  0.94    \\
DENIS-P J0250072$-$860930 & 02 50 07.20     & $-$86 09 30.0  & 1999.712 & ~\,9.26 &  2.09   &  1.41    \\
DENIS-P J0441247$-$271453 & 04 41 24.70     & $-$27 14 53.6  & 1999.063 & ~\,8.92 &  2.19   &  1.20    \\ 
DENIS-P J1117420$-$264453 & 11 17 42.08     & $-$26 44 53.8& 1999.266 &  11.71  &  2.16   &  0.56    \\ 
DENIS-P J1236396$-$172216 & 12 36 39.61     & $-$17 22 16.9  & 1999.384 &  13.91  &  2.14   &  1.14    \\ 
DENIS-P J1400335$-$271656 & 14 00 33.51     & $-$27 16 56.2  & 1999.348 & ~\,9.69 &  2.09   &  1.26    \\
DENIS-P J1405376$-$221515 & 14 05 37.64     & $-$22 15 15.0  & 1999.285 & ~\,9.49 &  2.09   &  1.29    \\
DENIS-P J1427297$-$264040 & 14 27 29.71     & $-$26 40 40.8  & 1999.419 & ~\,9.68 &  2.12   &  1.20    \\
DENIS-P J1510397$-$212524 & 15 10 39.72     & $-$21 25 24.9  & 1999.384 &  10.06  &  2.22   &  1.18    \\
DENIS-P J1525014$-$032359 & 15 25 01.46     & $-$03 23 59.5 &1999.351 & ~\,9.25 &  2.09   &  1.08    \\
DENIS-P J1552237$-$033520 & 15 52 23.78     & $-$03 35 20.7 &1999.534 &  12.02  &  2.07   &  1.37    \\
DENIS-P J1552551$-$045215 & 15 52 55.19     & $-$04 52 15.3 &1999.534 &  10.21  &  2.01   &  1.38    \\
DENIS-P J1553186$-$025919 & 15 53 18.65     & $-$02 59 19.3   &1999.581 &  13.12  &  2.04   &  1.36    \\
DENIS-P J1615446$-$040526 & 16 15 44.69     & $-$04 05 26.2 &1999.353 & ~\,9.67 &  2.03   &  1.19    \\
DENIS-P J1618120$-$044221 & 16 18 12.09     & $-$04 42 21.8 &1999.386 &  11.32  &  2.26   &  1.28    \\
DENIS-P J2024329$-$294402 & 20 24 32.96     & $-$29 44 02.6 &1999.392 &  10.45  &  2.13   &  1.26    \\
DENIS-P J2032270$-$273058 & 20 32 27.03     & $-$27 30 58.4   &1999.534 &  10.76  &  2.45   &  1.18    \\
DENIS-P J2034203$-$263652 & 20 34 20.33     & $-$26 36 52.2   &1999.712 &  11.27  &  2.23   &  1.12    \\
DENIS-P J2044066$-$173457 & 20 44 06.68     & $-$17 34 57.3 &1999.606 &  11.28  &  2.42   &  1.29    \\
DENIS-P J2055240$-$322600 & 20 55 24.07     & $-$32 26 00.8 &1999.669 &  10.73  &  2.10   &  1.30    \\  
DENIS-P J2056329$-$782540 & 20 56 32.90     & $-$78 25 40.1   &1999.660 &  10.43  &  2.08   &  1.20    \\
DENIS-P J2058075$-$730350 & 20 58 07.55     & $-$73 03 50.4   &1999.660 &  11.89  &  2.35   &  1.29    \\
DENIS-P J2103375$-$783831 & 21 03 37.56     & $-$78 38 31.5 &1999.658 &  11.42  &  2.08   &  1.30    \\
DENIS-P J2124575$-$341655 & 21 24 57.51     & $-$34 16 55.9   &1999.559 &  13.60  &  2.37   &  1.32    \\
DENIS-P J2125399$-$100526 & 21 25 39.98     & $-$10 05 26.1 &1999.482 &  12.25  &  2.10   &  0.59    \\
DENIS-P J2130021$-$815158 & 21 30 02.15     & $-$81 51 58.6   &1999.510 &  10.33  &  2.17   &  1.37    \\
DENIS-P J2155040$-$165535 & 21 55 04.06     & $-$16 55 35.1 &1999.712 &  12.46  &  2.37   &  0.40    \\
DENIS-P J2203522$-$593300 & 22 03 52.29     & $-$59 33 00.7 &1999.649 &  11.29  &  2.43   &  1.16    \\ 
DENIS-P J2206227$-$204706 & 22 06 22.78     & $-$20 47 06.0 &1999.611 &  15.09  &  2.67   &  1.22    \\
DENIS-P J2225004$-$121606 & 22 25 00.48     & $-$12 16 06.9 &1999.447 &  10.38  &  2.25   &  1.19    \\
DENIS-P J2226443$-$750342 & 22 26 44.36     & $-$75 03 42.7 &1999.814 &  15.20  &  2.84   &  1.20    \\ 
DENIS-P J2312219$-$091513 & 23 12 21.98     & $-$09 15 13.5 &1999.482 &  13.70  &  2.99   &  0.29    \\ 
DENIS-P J2334544$-$193232 & 23 34 54.49     & $-$19 32 32.4 &1999.707 &  14.63  &  2.94   &  0.54    \\
    \noalign{\smallskip}
    \hline 
   \end{tabular}
  $$
  \begin{list}{}{}
   \item[$ $]
   
Columns 1: Object name. 

Columns 2, 3 \& 4: DENIS Position with respect to equinox J2000 at DENIS epoch.

Columns 5, 6 \& 7: DENIS I-magnitude and colours.
  \end{list}
\end{table*}
Color-magnitude relations have yet to be established for the native
DENIS photometric system, and we therefore resort to a relation in the
standard Cousins-CIT system to derive distances for red stars in the 
DENIS database. A preliminary comparison of the two photometric systems 
for very late M dwarfs shows differences of $\sim$0.1~mag between the 
K bands, of slightly less than 0.05~mag between the J bands, and below 
0.05~mag between the two I bands (Delfosse \cite{delfosseb}).
We therefore choose to determine distances from the (I$-$J~,~M$_{\rm I}$) 
relation, which is least affected by the neglected colour terms, and
note that systematic errors from adopting the Cousins-CIT relation are much
smaller than the intrinsic dispersion of that relation.
The two analytic (I$-$J~,~M$_{\rm I}$) relations in the literature, from Leggett 
(\cite{leggett}) and Tinney (\cite{tinney93}), respectively concentrate
on early and late M dwarfs.
For consistency across the M spectral class (and for convenience), we 
therefore fitted a polynomial relation to the photometric 
and trigonometric data of Leggett (\cite{leggett}), Tinney
(\cite{tinney93}) and Tinney et al. (\cite{tinney}), adding some 
photometry from Alonso et al. (\cite{alonso}) and Weis
(\cite{weis}). Figure~\ref{fig_col_mag} shows the resulting relation, 
\begin{eqnarray}
M_{\rm I} & = & 4.97-4.76({\rm I-J})+10.03({\rm I-J})^{\rm 2}-4.07({\rm I-J})^{\rm 3} \nonumber \\
      &   &+0.53({\rm I-J})^{\rm 4} \label{eq1},
\end{eqnarray}
valid for $1.0<$I$-$J$<3.0$

and demonstrates its fair agreement with the theoretical track of Baraffe et
al. (\cite{baraffe98}). The new polynomial fit complements the
(I$-$K~,~M$_{\rm K}$) relation derived by Tinney et al. (\cite{tinney}) 
for the same spectral type range.

With the above Color-Magnitude relation as tool, we have searched the 
first 2110 squares degrees processed 
by PDAC (Figure~\ref{ij_jk}) for new members of the solar neighbourhood. 
We select all DENIS sources with $|b_{\rm II}| \geq 30\degr$ that have the 
I$-$J colour of an M dwarf ($1.0 \leq $I$-$J$ \leq 3.0$; Leggett \cite{leggett}). 
We then compute their photometric distance (D$_{\rm phot}$) from the 
DENIS photometry and the above polynomial colour-magnitude relations,
and retain those with D$_{\rm phot} \leq 30$~pc.
In the present paper we focus on the 2.0~$<$~I$-$J~$<$~3.0 range, or 
spectral types of approximately M6 to M8. Bluer nearby M dwarfs are brighter and hence
less likely to have been overlooked. At present we therefore give them
a lower priority. Redder ones, like DENIS-P~J104814.7-395606.1 (Delfosse et
al. \cite{delfossec}), are addressed by the DENIS Brown Dwarf program (Delfosse et al., 
in preparation).

These criteria are met by 60~objects
(Table~\ref{obs},~\ref{obs1}), 13 of which are listed in proper motion 
catalogues but had no distance estimate (except WT~792, discussed below).
At that stage the sample is a mix of
distant giants and nearby dwarfs, which cannot be completely separated
using the DENIS photometry alone: the dwarf and giant sequences 
are distinct in the DENIS I$-$J/J$-$K colour-colour diagram, but 
not sufficiently separated to distinguish the two classes with 100\% 
completeness and reliability.

\subsection{Additional candidates from the LHS catalogue}

We additionally browsed the LHS catalogue (Luyten \cite{luytena}) for
objects that match the colour and photometric distance criteria, but which
had been missed because of our galactic latitude cutoff, or whose DENIS 
observations had not yet been entered in the database. To narrow down 
that search, we looked for unambiguous DENIS identifications to southern 
LHS stars matching B$_{\rm Luyten}$~$>$~16.0, R$_{\rm Luyten}$~$>$~15.0, and 
B$_{\rm Luyten}-$R$_{\rm Luyten}$~$>$~1.5, and then applied the colour and 
photometric distance cutoffs. This identified 4 additional candidates: 
LHS 1810, 2049, 2859 and 5165, which are added in the Tables.

\section{Proper motions and B, R photometry}

\begin{table}
   \caption{Proper motion of the 30 Nearby Star candidates}
    \label{pm}
  $$
   \begin{tabular}{lllll}
   \hline 
   \hline
   \noalign{\smallskip}
        DENIS Name  & $\mu_{\rm \alpha}$ & $\mu_{\rm \delta}$ & $\mu_{\rm total}$ & $\mu_{\rm _L/W}$ \\ 
                    & [\arcsec yr$^{\rm -1}$]   & [\arcsec yr$^{\rm -1}$]   &[\arcsec yr$^{\rm -1}$]   & [\arcsec yr$^{\rm -1}$] \\ 
           \hline
           \noalign{\smallskip}
        J0004575$-$170937*& +0.146 & $-$0.011 & 0.146 & ...  \\
        J0013466$-$045736 & +0.584 & $-$0.153 & 0.604 &0.619 \\        
        J0019275$-$362015*& +0.152 & $-$0.097 & 0.180 & ... \\
        J0041353$-$562112*& +0.121 & $-$0.064 & 0.137 & ...  \\
        J0145434$-$372959*& +0.424 & $-$0.118 & 0.440 & ... \\
        J0253444$-$795913*& +0.074 & +0.081 & 0.110 & ... \\
        J0312251$+$002158*& +0.173 & $-$0.028 & 0.176 & ...  \\
        J0324268$-$772705*& +0.287 & +0.184 & 0.341 & ... \\ 
        J0413398$-$270428 & +0.234 & $-$0.023 & 0.235 & 0.230 \\
        J0602542$-$091503 & +0.172 & $-$0.616 & 0.639 & 0.608 \\
        J0848189$-$201911 & +0.342 & $-$0.593 & 0.685 & 0.633 \\
        J1003191$-$010507 & $-$0.497 & +0.056 & 0.500 & 0.491 \\
        J1136409$-$075511 & $-0$.171 & +0.144 & 0.224 & 0.223 \\
        J1216101$-$112609 & +0.041 & $-$0.219 & 0.223 & 0.212 \\
        J1223562$-$275746 & $-$1.239 & +0.323 & 1.280 & 1.293 \\
        J1236153$-$310646 & +0.161 & $-$0.083 & 0.181 & 0.191 \\
        J1357149$-$143852*& $-$0.355 & +0.023 & 0.356 & ... \\
        J1406493$-$301828 & $-$0.855 & $-$0.059 & 0.857 & 0.814 \\
        J1412069$-$041348 & +0.292 & $-$0.152 & 0.329 & 0.342 \\
        J1553251$-$044741*& $-$0.034 & $-$0.121 & 0.126 & ... \\ 
        J1614252$-$025100 & +0.003 & +0.367 & 0.367 & 0.401 \\ 
        J2002134$-$542555*& +0.060 & $-$0.364 & 0.369 & ... \\ 
        J2049527$-$171608 & +0.303 & $-$0.102 & 0.320 & 0.305 \\ 
        J2107247$-$335733*& +0.344 & $-$0.367 & 0.503 & ... \\ 
        J2134222$-$431610 & +0.142 & $-$0.792 & 0.804 & 0.785 \\
        J2202112$-$110945 & +0.137 & $-$0.182 & 0.228 & 0.210 \\
        J2213504$-$634210 & +0.140 & +0.166 & 0.217 & 0.195 \\
        J2331217$-$274949*& +0.087 & +0.738 & 0.744 & ... \\ 
        J2333405$-$213353 & +0.682 & $-$0.336 & 0.761 & 0.794  \\ 
        J2353594$-$083331*& $-$0.039 & $-$0.367 & 0.369 & ... \\ 
    \noalign{\smallskip}
    \hline 
   \end{tabular}
  $$
  \begin{list}{}{}
  \item[$^{*}$] Not previously known as a high proper motion star
    
  Columns 2, 3 \& 4: $\mu_{\rm \alpha}$, $\mu_{\rm \delta}$, $\mu_{\rm total}$ 
    our measurement, in arc-sec.yr$^{\rm -1}$. 
  
  Columns 5: Literature total proper motion, from Luyten (\cite{luytena}, 
   \cite{luytenb}) or Wroblewski et al. 
  (\cite{wroblewski}) when available.
  \end{list}
\end{table}

As an initial step towards rejecting
giants, we determine the proper motion of the candidates from a comparison
of the DENIS position with archival Schmidt plates digitized on the MAMA 
microdensitometer. Objects fainter than R~=~10 with a proper motion 
$\mu \geq 0.1$\arcsec yr$^{\rm -1}$ must be dwarfs: a red giant with such 
an apparent magnitude and proper motion would have a very large and improbable 
space velocity ($\ge 1000$ km/s), much larger than the Galactic escape speed 
(Meillon \cite{meillon}). Conversely, some small fraction of the
nearby dwarfs must have proper motions below our cutoff.
\begin{figure}
\psfig{height=6.5cm,file=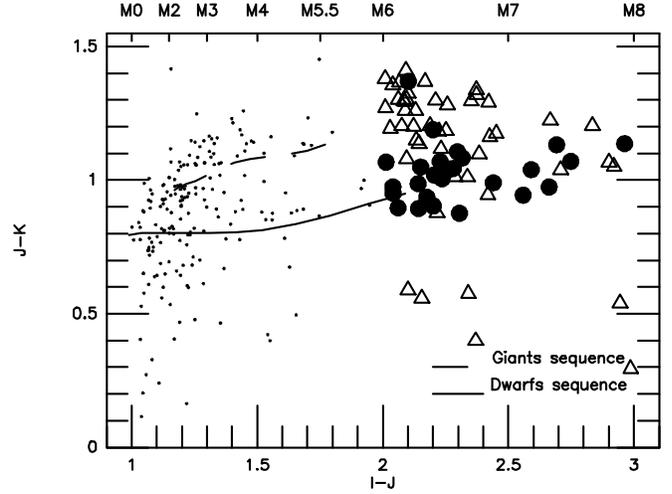,angle=-90} 
\caption{All DENIS sources with the I$-$J colour of an M-dwarf 
(1.0~$<$~I$-$J~$<$~3.0) detected in the first 2110 squares degrees 
processed by PDAC, with a photometric distance $\leq$ 30~pc and a
galactic latitude ${\geq}~30\degr$. The dots represent the early
M-stars (1.0~$<$~I$-$J~$<$~3.0). Amongst the 60 DENIS sources 
in the 2.0$-$3.0 range (corresponding to a late-M spectral class) 
the solid circles represent the objects with a proper motion 
$\mu$~$>$~0.1\arcsec yr$^{\rm -1}$, and the triangles those with a smaller proper 
motion. The solid and dashed lines
respectively represent the dwarf and giant sequences of Bessell \& Brett
(1988), for slightly different filters. The (indicative) spectral 
type labels on the top axis are adopted from Leggett \cite{leggett}.
}
\label{ij_jk}
\end{figure}
To estimate that proportion, we counted the fraction of the stars within
25~pc in both the CNS3 and the Hipparcos catalogue that have a proper
motion smaller than 0.1\arcsec yr$^{\rm -1}$. In the CNS3, this fraction of "slow" stars 
is about 13\% for limiting magnitudes of V~$<$~7, 8 and 9. In the Hipparcos 
catalogue and for the same limiting magnitudes it is much lower,
only 6.5\%. The CNS3 is considered complete to V~=~9. The Hipparcos 
catalogue is complete up to V~=~7.3, and nearly
complete up to at least 9 for nearby stars, as potential nearby 
stars were systematically included in the Hipparcos Input catalogue.
Incompleteness therefore cannot explain the discrepancy between the
fractions of low proper-motion stars in the two catalogues.
A careful comparison between the two lists of "slow"
stars shows that half of the slow CNS3 objects have a Hipparcos
distance larger than 25 pc, and should therefore not be taken into
consideration. The fraction of nearby stars lost due to our proper motion 
cut-off is therefore of the order of 6\%. Our future 
goal is to obtain spectra of
all candidates, independently of their proper motions, to assemble
a complete and unbiased inventory of the solar neighbourhood for the 
high galactic latitude southern sky.

To obtain B and R photometry of the candidates and determine their proper 
motions, we identified all Survey plates in the CAI 
{\footnotesize (http://dsmama.obspm.fr/)} plate vault that
contained images of the nearby star candidates. Depending on the  
declination of the candidates, the following plates were available: POSS I for 
$-30\degr<\delta<0\degr$, SRC-J for $-90\degr<\delta<0\degr$, 
SRC-R for 
$-17\degr<\delta<0\degr$, and ESO-R for $\delta<-17\degr$. 
We used the MAMA microdensitometer (Berger et al. \cite{berger}) at 
CAI to digitize the survey plates,
and analysed the resulting images with SExtractor (Bertin \& Arnouts 
\cite{bertin}). 
\begin{figure*}
\resizebox{\hsize}{!}{\includegraphics{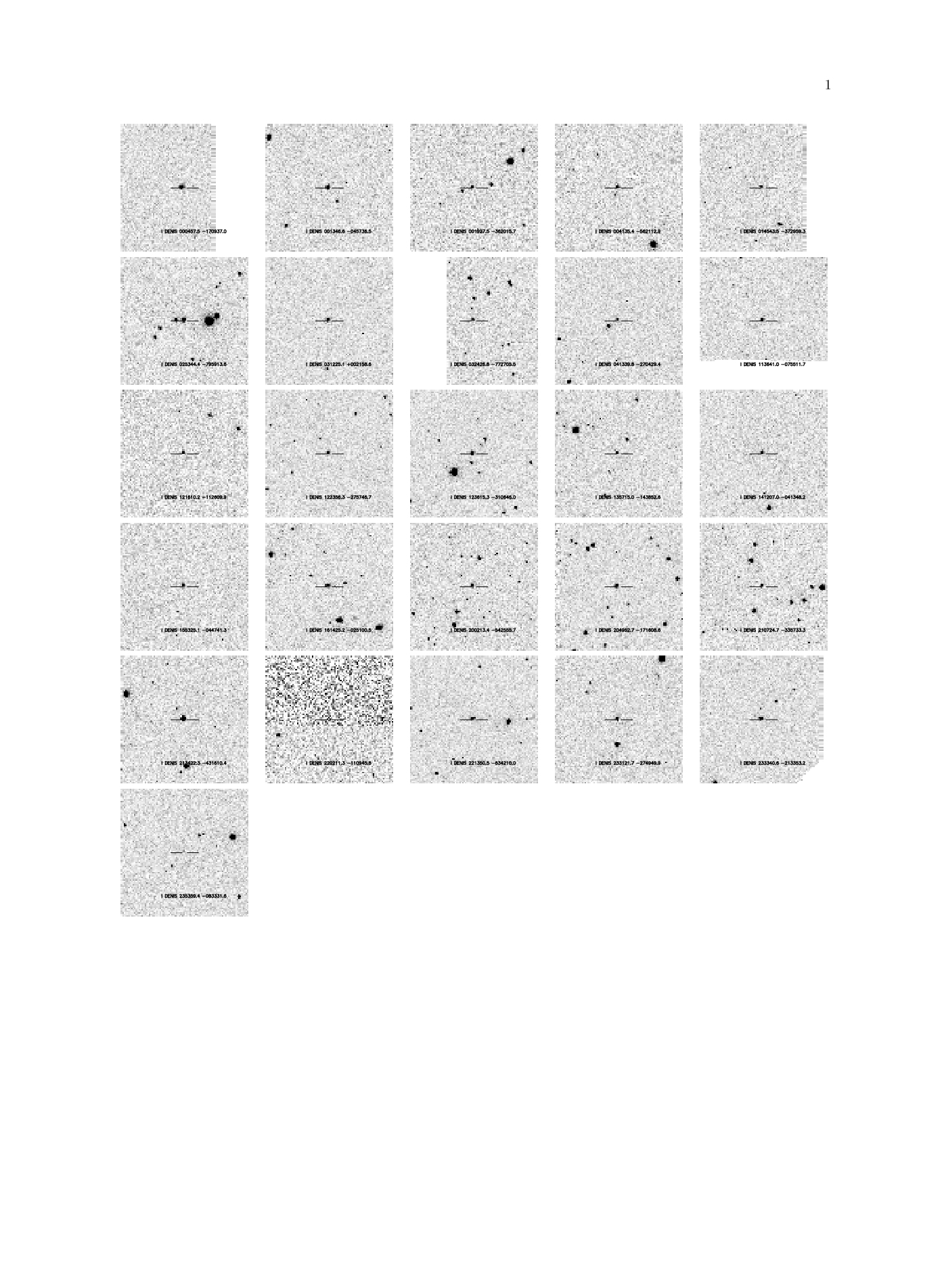}}
\caption{I-band finding charts for the objects listed in Table~\ref{obs}.
The size of each chart is $\sim~3.5'~\times~3.5'$. North is up and east is to
the left. The 4 stars selected directly from the LHS are not included.}
\label{charts}
\end{figure*}
The SExtractor source parameters were calibrated using the 
GSPC-2 (Postman et al. \cite{postman}, Bucciarelli et al. 
\cite{bucciarelli}) and ACT catalogues (Urban et al. \cite{urban}) as 
photometric and astrometric references to produce B and R magnitudes 
(Table~\ref{dis}),
and equatorial coordinates at the epoch of the plate. Absolute proper 
motions were then determined through a least square fit to the positions
at the 2 to 4 Schmidt plate epochs and the DENIS survey epoch. 
The proper motion errors range between 8 and 19 mas/year, depending on the
object magnitude and the available time baseline. The magnitude errors are 
$\pm 0.3$ mag for B and $\pm 0.2$ mag for R. 
\begin{table*}
   \caption{MAMA and DENIS photometry, and estimated distances, for the 
   30 DENIS red dwarf candidates}
    \label{dis}
  $$
   \begin{tabular}{lllllllllll}
   \hline
   \hline
   \noalign{\smallskip}
DENIS objects     & B   &   R  & I     & I$-$J  & J$-$K & M$_{\rm I}$&M$_{\rm K}$& Dist$_{\rm I}$ &Dist$_{\rm K}$ & V$_{\rm t}$ \\ 
                  &     &      &       &        &       &        &       & [pc]       & [pc]      & [km/s]  \\ 
   \hline
   \noalign{\smallskip}
J0004575$-$170937 & 18.2& 15.6 & 13.00 & 2.03 & 0.93  &  11.59 &~\,8.78 & 19.1 & 17.8  &~\,13.2  \\   
J0013466$-$045736 & 19.5& 16.8 & 13.88 & 2.44 & 0.99  & 12.73  &~\,9.69 & 17.0 & 14.2  &~\,48.7  \\        
J0019275$-$362015 & 18.2& 17.0 & 14.30 & 2.14 & 0.89  & 11.95  &~\,8.95 & 29.6 & 29.1  &~\,25.3  \\
J0041353$-$562112 & 20.6& 18.7 & 14.68 & 2.74 & 1.07  & 13.37  & 10.12  & 18.2 & 14.1  &~\,11.8  \\
J0145434$-$372959 & 20.2& 17.7 & 15.05 & 2.56 & 0.94  & 13.00  &~\,9.79 & 25.7 & 22.5  &~\,53.6  \\
J0253444$-$795913 & 18.7& 16.2 & 13.47 & 2.15 & 0.98  & 11.98  &~\,9.17 & 19.9 & 17.1  &~\,10.4  \\ 
J0312251$+$002158 & 20.4& 17.0 & 14.30 & 2.17 & 0.93  & 12.03  &~\,9.11 & 28.4 & 26.2  &~\,23.7  \\ 
J0324268$-$772705 & 20.4& 17.3 & 14.36 & 2.26 & 1.04  & 12.29  &~\,9.49 & 26.0 & 20.6  &~\,42.0  \\
J0413398$-$270428 & 22.0& 17.7 & 14.45 & 2.26 & 0.99  & 12.29  &~\,9.40 & 27.1 & 22.9  &~\,30.2  \\
J0602542$-$091503 & 17.3& 15.6 & 12.98 & 2.12 & 0.85  & 11.88  &~\,8.81 & 16.6 & 17.4  &~\,48.9  \\
J0848189$-$201911 & 19.7& 17.2 & 14.51 & 2.21 & 0.98  & 12.15  &~\,9.29 & 29.7 & 25.4  &~\,88.3  \\
J1003191$-$010507 & 21.0& 18.0 & 14.94 & 2.62 & 1.08  & 13.12  & 10.02  & 23.1 & 17.6  &~\,51.7  \\
J1136409$-$075511 & 17.8& 16.5 & 14.31 & 2.20 & 0.90  & 12.12  &~\,9.11 & 27.4 & 26.3  &~\,29.1  \\
J1216101$-$112609 & 19.5& 17.1 & 14.72 & 2.30 & 1.10  & 12.39  &~\,9.65 & 29.2 & 21.6  &~\,30.9  \\
J1223562$-$275746 & 19.6& 16.4 & 14.19 & 2.30 & 0.88  & 12.39  &~\,9.27 & 22.9 & 22.3  & 138.9  \\
J1236153$-$310646 & 18.5& 16.7 & 13.97 & 2.23 & 1.01  & 12.21  &~\,9.39 & 22.5 & 18.6  &~\,19.3  \\
J1357149$-$143852 & 21.1& 18.5 & 15.55 & 2.69 & 1.14  & 13.27  & 10.13  & 28.5 & 20.8  &~\,48.1  \\
J1406493$-$301828 & 18.5& 15.7 & 13.42 & 2.13 & 1.01  & 11.91  &~\,9.19 & 20.0 & 16.5  &~\,77.5  \\
J1412069$-$041348 & 19.1& 16.3 & 13.66 & 2.04 & 0.98  & 11.63  &~\,8.93 & 25.5 & 22.0  &~\,39.8  \\
J1553251$-$044741 & 17.4& 15.1 & 13.67 & 2.10 & 1.37  & 11.82  &~\,9.75 & 23.4 & 12.3  &~\,14.0  \\
J1614252$-$025100 & 19.3& 17.3 & 13.63 & 2.29 & 1.04  & 12.37  &~\,9.54 & 17.9 & 14.2  &~\,31.1  \\ 
J2002134$-$542555 & 19.4& 17.1 & 13.89 & 2.20 & 1.18  & 12.12  &~\,9.62 & 22.6 & 15.1  &~\,39.5  \\
J2049527$-$171608 & 18.9& 16.8 & 14.16 & 2.32 & 1.08  & 12.44  &~\,9.65 & 22.0 & 16.7  &~\,33.4  \\ 
J2107247$-$335733 & 19.9& 16.7 & 14.36 & 2.20 & 1.02  & 12.12  &~\,9.35 & 28.0 & 22.8  &~\,66.8  \\
J2134222$-$431610 & 17.6& 15.3 & 12.78 & 2.02 & 1.06  & 11.56  &~\,9.07 & 17.5 & 13.4  &~\,66.7  \\
J2202112$-$110945 & 20.7& 17.9 & 15.11 & 2.66 & 0.98  & 13.21  &~\,9.96 & 24.0 & 20.1  &~\,25.9  \\
J2213504$-$634210 & 18.7& 16.3 & 13.05 & 2.16 & 1.04  & 12.01  &~\,9.31 & 16.2 & 12.8  &~\,16.7  \\
J2331217$-$274949 & 20.4& 17.7 & 14.25 & 2.59 & 1.04  & 13.06  &~\,9.94 & 17.3 & 13.6  &~\,61.0  \\
J2333405$-$213353 & 19.3& 16.5 & 13.89 & 2.06 & 0.90  & 11.69  &~\,8.78 & 27.5 & 26.9  &~\,98.2  \\ 
J2353594$-$083331 & **  & 18.6 & 15.93 & 2.97 & 1.13  & 13.92  & 10.32  & 25.3 & 20.0  &~\,44.3  \\ 
   \noalign{\smallskip}
   \hline
   \end{tabular}
  $$
  \begin{list}{}{}
  \item[$^{**}$] Too faint for the plate   
      
  Column 1: Object name.     
  Columns 2, 3: B, R magnitudes determined from the plates using GSPC and GSPC-2 calibrations.     
  Columns 4, 5, 6: DENIS I-magnitude and colours.    
  Column 7: M$_{\rm I}$ absolute I band magnitude, from the colour magnitude 
   relation derived in the present paper.     
  Column 8: M$_{\rm K}$ absolute K band magnitude, from the Tinney et al. 
   (\cite{tinney}) relation.     
  Columns 9, 10: Dist$_{\rm I}$, Dist$_{\rm K}$ distances estimated from 
    M$_{\rm I}$ and M$_{\rm K}$ respectively.      
  Column 11: V$_{\rm t}$ tangential velocity, computed from Dist$_{\rm I}$ and the 
   proper motion.   
  \end{list}
 \end{table*}
>From the full sample of 60 objects, 34 have no proper motion above 
the $0.1$\arcsec yr$^{\rm -1}$ level and are listed in Table~\ref{obs1}. They are
not studied in this paper, and leave us with 26 high quality
nearby star candidates (Table~\ref{obs}) to which the 4 LHS stars are added
(finding charts in Fig.~\ref{charts}).
Of the 30 proper-motion objects, 13 were previously unknown.
2 of those have a proper motion, $\mu > 0.5$\arcsec yr$^{\rm -1}$, that would have
qualified them
for inclusion in the LHS. An additional 5 have $\mu > 0.2$\arcsec yr$^{\rm -1}$
and could have been included in the NLTT. The 17 remaining objects 
belong to the LHS, the NLTT, or the Catalogue of 
New proper motion stars (Wroblewski \& Torres \cite{wroblewski}). 
For those, the difference between our 
proper motions and the literature values has a mean of 9~mas\,yr$^{\rm -1}$ 
and a dispersion 
of 23~mas\,yr$^{\rm -1}$, of the order of the quoted LHS accuracy.
Table~\ref{pm} lists our proper motion measurements, and the
literature proper motion when available.
Figure~\ref{ij_jk}, which distinguishes the high and low proper motion 
stars in
the I$-$J/J$-$K diagram, immediately shows that the high proper 
motion stars belong to the dwarf sequence, with only one exception,
DENIS-P J1553251-044741.
On the other hand, some low proper motion stars do, as expected, seem 
to belong to the dwarf sequence.

\section{Discussion}  

In addition to the (I~,~I$-$J) distances computed from the relation derived
above, Table~\ref{dis} contains (I~,~I$-$K) distances derived using 
the (I$-$K~,~M$_{\rm K}$) colour-magnitude relation of Tinney et al. 
(\cite{tinney}) for $1.2<$I$-$K$<5.4$. The latter are systematically somewhat smaller, 
most likely because of our use of DENIS K$_s$ magnitudes and a
colour-magnitude relation established in the Cousins-CIT photometric
system. The (I~,~I$-$J) distances are therefore most likely more reliable.
Carpenter (\cite{carpenter}) carefully derived relations for colour transformations
between 2MASS and other systems, including CIT and DENIS.
However the DENIS data he used is the part publicly released at CDS and reduced by
the Leiden Data Center (LDAC), and covers only a very small fraction of the DENIS data.
The data presented here was reduced by the Paris Data Center (PDAC) which uses
slightly different methods and relations. Therefore the Carpenter's equations
should not be used here. 
We note that Patterson et al. (\cite{patterson}) estimate a distance
of 22~pc to WT~792 from VRI photometry that is compatible with our
Dist$_{\rm I}$ within the errors.

It must be realized that the distances in Table~\ref{dis} carry considerable 
uncertainties, as any photometric parallaxes do. The intrinsic scatter 
of the colour-luminosity relation (Fig.~\ref{fig_col_mag}) is $\pm$1~magnitude, 
which on a star by star basis translates into a 45\% random distance 
error. At a lower level, the poynomial relations most likely
have systematic errors at some colours: 0.3~magnitude local 
errors (i.e. 14\% distance errors) can quite easily creep in when 
fitting a phenomenological model to such a dispersed diagram.

Finally, some of the 30~objects must be multiple systems,
whose distances are underestimated by attributing their
luminosity to one single star.

Tangential velocities calculated from the proper motions and distances,
are listed in the last column of Table~\ref{dis}; 4 stars have tangential 
velocities larger than 70 km/s: LHS 2049; 325a; 2859; 3970. Of those
LHS~325a and LHS~2049 were previously recognized as late-M stars with 
high reduced proper motion (H = m + 5 log $\mu$ + 5) by Bessell 
(\cite{bessell}). LHS~325a is photometrically classified as an old-disk 
star by Leggett (\cite{leggett}).

The sample of low proper motion objects (Table~\ref{obs1}) probably contains a few 
nearby stars, that happen to have a low tangential velocity, and is presently
under study.
   
The I$-$J and I$-$K colours of the 30 nearby star candidates indicate
spectral types of M6 to M8, which we plan to ascertain through 
low-resolution spectroscopy. Such objects have masses of 0.1M$_{\rm \odot}$
(Delfosse et al. \cite{delfosse00}) or lower.

The 2110 square degrees explored here contain 26 stars with 
2.0~$<$~I$-$J~$<$~3.0, 
$\mu$~$>$~0.1\arcsec yr$^{\rm -1}$, and a photometric distance d~$<$~30~pc. The whole 
celestial sphere therefore contains $\sim$~500 similar objects, of which
$\sim$~60\%, 
or 300, will have a photometric distance below 25~pc. Malmquist bias from
our approximate photometric distances must bias this number up, and
the number of stars that are actually within 25~pc must be significantly 
smaller, perhaps by as much as a factor of 2. 
The (R$-$I)$_{\rm C}$ colour index that corresponds to our I$-$J~$=$~2.0 cutoff is
(R$-$I)$_{\rm C}$~$=$~2.0 (Leggett \cite{leggett}), or about 1.6 in the Kron system
used in the CNS3. 
The whole CNS3 contains only 23 stars with (R$-$I)$_{\rm Kron}$~$>$~1.6.
An effort similar to ours over the whole 
celestial sphere would thus identify 150 to 250 
new stars within 25~pc.

\section{Conclusion}

Many new nearby stars have been recently identified thanks to 
the near-IR sky surveys: DENIS and 2MASS have proved to be valuable
resources for the identification of new nearby stars. In the present
paper, we identify 60 photometric nearby star candidates in 2110 square 
degrees, with distances that lie between 15 and 30~pc.
26 of those have large proper motions that exclude that they are distant 
giants, 13 of which are new; 4 additional known LHS stars were
recognized as nearby star candidates outside the main search area. 
A few stars have large tangential
velocities and may be subdwarfs, whose distances would then be 
slightly overestimated. 
We plan to obtain low-resolution spectra for all candidates in the near
future.

\begin{acknowledgements}
We are grateful to the DENIS consortium for providing the DENIS data, and 
to Ren\'e Chesnel for scanning and 
pre-reducing the photographic plates. The long-term loan of POSS~I plates 
by the Leiden Observatory to Observatoire de Paris is gratefully acknowledged.

We thank the referee, Dr. Hartmut Jahrei\ss, for his prompt and very 
constructive report.
\end{acknowledgements}


\end{document}